\documentclass[12pt,a4]{article}

\usepackage{amsmath} 
\usepackage{graphicx} 
\usepackage{floatflt} 
\usepackage{subfigure}
\usepackage{theorem} 
\usepackage{url}
\usepackage{algorithmic}
\newcommand{\tr}{\mathrm{tr}} 
\newcommand{\bra}[1]{\langle#1|}

\newcommand{\ket}[1]{|#1\rangle} 
\newcommand{\braket}[2]{\langle#1|#2\rangle}
\newcommand{\ketbra}[2]{|#1\rangle\langle#2|} 
\newcommand{\interm}[0]{|X|^2}
  
\newcommand{\poit}[2]{p(#1;#2)}

\newcommand{\amp}[2]{a(#1;#2)}
\newtheorem{lemma}{Lemma}
\newtheorem{theorem}{Theorem} 

\newtheorem{proposition}{Proposition} 
\newtheorem{definition}{Definition}

\newcommand{\qed}{\hfill \mbox{\raggedright \rule{.07in}{.1in}}}
\newenvironment{proof}{\vspace{1ex}\noindent{\bf Proof}\hspace{0.5em}}
{\hfill\qed\vspace{1ex}\linebreak}

\sloppypar
\begin{document}
\pagestyle{empty}
\title{Getting Beyond the State of the Art of Information Retrieval with Quantum Theory}
\author{Massimo Melucci\\\normalsize{University of Padua}}
\date{}
\maketitle
\begin{abstract}
  According to the probability ranking principle, the document set with the highest values of
  probability of relevance optimizes information retrieval effectiveness given the
  probabilities are estimated as accurately as possible. The key point of this principle is
  the separation of the document set into two subsets with a given level of fallout and with
  the highest recall. If subsets of set measures are replaced by subspaces and space measures,
  we obtain an alternative theory stemming from Quantum Theory. That theory is named after
  vector probability because vectors represent event like sets do in classical
  probability. The paper shows that the separation into vector subspaces is more effective
  than the separation into subsets with the same available evidence. The result is proved
  mathematically and verified experimentally. In general, the paper suggests that quantum
  theory is not only a source of rhetoric inspiration, but is a sufficient condition to
  improve retrieval effectiveness in a principled way.
\end{abstract}
\thispagestyle{empty}

\section{Introduction}
\label{sec:introduction}

Information Retrieval (IR) systems decide about the relevance under conditions of uncertainty.
As a measure of uncertainty is necessary, a probability theory defines the event space and the
probability distribution. The research in probabilistic IR is based on the classical theory of
probability, which describes events and probability distributions using, respectively, sets
and set measures obeying the usual axioms stated in~\cite{Kolmogorov56}.  Set theory is not
the unique way to define probability though.

If subsets and set measures are replaced by vector subspaces and space-based measures, we
obtain an alternative theory called, in this paper, \emph{vector probability}. Although this
theory stems from Quantum Theory, we prefer to use ``vector'' because vectors are sufficient
to represent events like sets represent events within classical probability, the latter being
the feature of our interest, whereas the ``quantumness'' of IR is out of the scope of this
paper, which explains that the replacement of classical with vector probability is crucial to
ranking.

Ranking is an essential task in IR. Indeed, it should not come as a surprise that the
Probability Ranking Principle (PRP) reported in~\cite{Robertson77} is by far the most
important theoretical result to date because it is an incisive factor in effectiveness.
Although probabilistic IR systems reach good results, ranking is far from being perfect
because irrelevant documents are often ranked at the top of, or useful units are missed from
the retrieved document list. 

Besides the definition of weighting schemes and ranking algorithms, new results can be
achieved if the research in IR views problems from a new theoretical perspective.  We propose
vector probability to describe the events and probabilities underlying an IR system. We show
that ranking in accordance with vector probability is more effective than ranking in
accordance with classical probability, given that the same evidence is available for
probability estimation. The effectiveness is measured in terms of probability of correct
decision or, equivalently, of probability of error. The result is proved mathematically and
verified experimentally.

Although the use of the mathematical apparatus of Quantum Theory is pivotal in showing the
superiority of vector probability (at least in terms of retrieval effectiveness), this paper
does not necessarily end in an investigation or assertion of quantum phenomena in IR.  Rather,
we argue that vector probability and then Quantum Theory is sufficient to go beyond the state
of the art of IR, thus supporting the hypothesis stated in~\cite{vanRijsbergen04} according to
which Quantum Theory may pave the way for a breakthrough in IR research.

\begin{figure*}[t]
  \centering
  \subfigure[][An IR system based on classical probability. A document is like an emitter of
  binary symbols referrint to term occurrence. The probability that $1$ occurs depends on a
  parameter that in turn depends on relevance. After training the probability distributions,
  the system B, which acts as a detector, decides whether a symbol is emitted by a relevant
  document. ]{
    \includegraphics[angle=270]{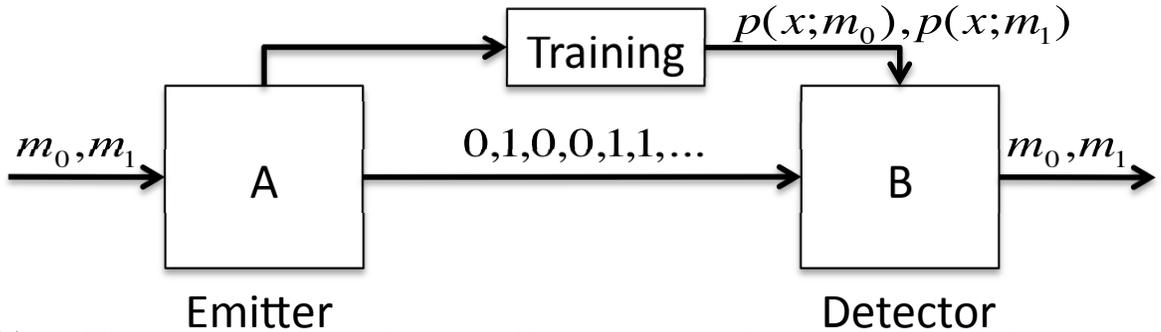}
    \label{ir-decision-a}}
  \subfigure[][An IR system equipped with an oracle based on vector probability works as the
  system of Figure~\ref{ir-decision-a} until symbols reaches an oracle which produces
  other symbols.  The symbols produced by the oracle are vectors. After training the
  probability distributions,  the system B, which acts as a detector, decides whether a vector
  is emitted by a relevant document.]{
    \includegraphics[angle=270]{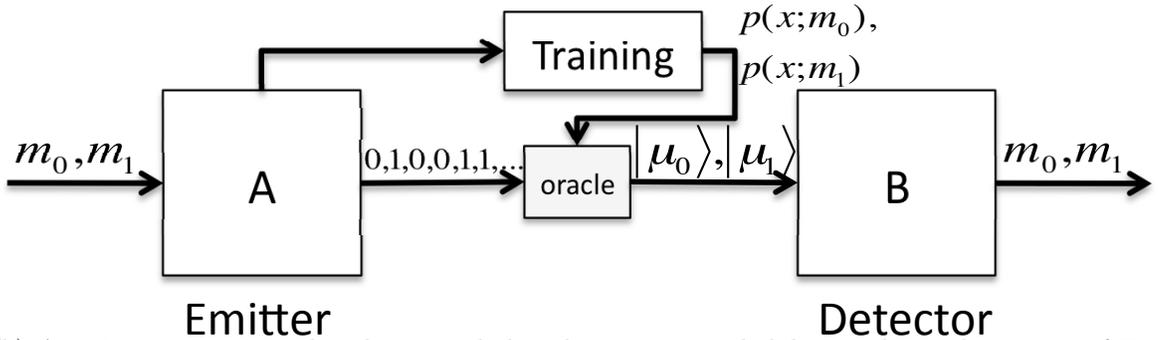}
    \label{ir-decision-b}}
  \caption{IR decision as an emitter and detector problem.}
  \label{fig:ir-decision}
\end{figure*}

We organize the paper as follows.  The paper gives an intuitive view of the our contribution
in Section~\ref{sec:intuitive-view} and sketches how an IR system built on the premise of
Quantum Theory can outperform any other system.  Section~\ref{sec:prob-relev} briefly reviews
the classical probability of relevance before introducing the notion of vector probability in
Section~\ref{sec:vect-prob}.  Section~\ref{sec:optim-observ} is one of the central sections of
the paper because it introduces the optimal vectors which are exploited in
Section~\ref{sec:vect-prob-relev-1} where we provide our main result, that is, the fact that a
system that ranks documents according to the probability of occurrence of the optimal vectors
in the documents is always superior to a system which ranks documents according to the
classical probability of relevance which is based on sets.
Section~\ref{sec:getting-beyond-state} addresses the case of BM25 and how it can be framed
within the theory. An experimental study is illustrated in Section~\ref{sec:experiments} for
measuring the degree to which vector probabilistic models outperforms classical probabilistic
models if a realistic test collectio is used.  The feasibility of the theory is strongly
dependent on the existence of an oracle which tells whether optimal vector occur in documents;
this issue is discussed in Section~\ref{sec:regarding-oracle}. After surveying the related
work in Section~\ref{sec:related-work}, we conclude with Section~\ref{sec:conclusions}.  The
appendix includes the definitions used in the paper and the proofs of the theoretical results.

\section{Intuitive View}
\label{sec:intuitive-view}

Before entering into mathematics, Figure~\ref{fig:ir-decision} depicts an intuitive view of
what is illustrated in the rest of the paper.  Suppose that relevance and non-relevance are
two events $m_0, m_1$ occurring with prior probability $1-\xi$ and $\xi$, respectively.
Document A is in either relevant or non-relevant.  Let's view A as an emitter of binary
symbols $0,1$ referring to presence or absence of a given index term. (We use the binary
symbol and relevance for the sake of clarity.)  On the other side, an IR system B acts as a
detector which has to decide whether a symbol comes out from either a relevant or a
non-relevant document.  The B's decision is taken on the basis of some feedback mechanism and
on the relevance and non-relevance probability distributions which has been appropriately
estimated on received symbols. An IR system that implements classical probability decides
about relevance without any transformation of the received symbols
(Figure~\ref{ir-decision-a}) whereas a IR system that implements vector probability decides
about relevance after a transformation of the received symbols carried out by an oracle which
outputs new symbols which cannot be straightforwardly derived from the received symbols
(Figure~\ref{ir-decision-b}) but can be defined as vectors~\cite{Helstrom76}. In this paper,
we show that when B is equipped with such an oracle, then it does significantly outperform any
other IR system which implements any classical probabilistic model. We theoretically measure
the improvement in effectiveness on the basis of a mathematical proposition which holds for
every IR system described in Figure~\ref{fig:ir-decision}.

\vfill
\section{Probability of Relevance}
\label{sec:prob-relev}

An IR system performing like detector B of Figure~\ref{ir-decision-a} computes the
probabilities that a symbol (e.g., an index term) occurs in relevant documents and in
non-relevant documents.  According to the intuitive view in terms of emitters and detectors
provided in Section~\ref{sec:intuitive-view}, the probability that a symbol (e.g., an index
term) occurs in relevant documents and in non-relevant documents is called, respectively,
\emph{probability of detection} ($P_d$) and \emph{probability of false alarm} ($P_0$). These
probabilities are also known as expected recall and fallout, respectively~\cite{Robertson77}.
The system decides whether a document is retrieved by
\begin{eqnarray}
  \frac{P_d(1-P_0)}{P_0(1-P_d)} &>& \lambda		\label{eq:2}
  % \\
  % \log{P_d(1-P_0)} - \log{P_0(1-P_d)} &>& \log\lambda	\label{eq:6}
  % \\
  % {P_d} - \lambda{P_0} &>& (1-P_d) - \lambda(1-P_0)      \label{eq:7}
\end{eqnarray}
where $\lambda$ is an appropriate threshold, and ranks the retrieved documents by using the
left side of~\eqref{eq:2}. For instance, when indipendent Bernoulli random variables are used,
we have that
\begin{equation*}
  P_d \propto \Pi_{j=1}^k p_j^{x_j} (1-p_j)^{1-x_j} \qquad P_0 \propto \Pi_{j=1}^k q_j^{x_j} (1-q_j)^{1-x_j}
\end{equation*}
where $p_j, q_j$ are the probabilities that term $j$ occurs in relevant, non-relevant
documents and the $x_j$'s belong to the region of acceptance. Depending on the available
evidence the probabilities are estimated as accurately as possible and are transformed into
weights (e.g., the binary weight or the BM25 illustrated in~\cite[page 340]{Robertson&09}).

The Probability Ranking Principle (PRP) defines the optimal document subsets in terms of
expected recall and fallout. Thus, the optimal document subsets are those maximizing
effectiveness. The PRP states that, if a cut-off is defined for expected fallout, that is,
probability of false alarm, we would maximize expected recall if we included in the retrieved
set those documents with the highest probability of relevance~\cite[page 297]{Robertson77},
that is, probability of detection. When a collection is indexed, each document belongs to
subsets labeled by the document index terms and the documents in a subset are
indistinguishable. In fact, \eqref{eq:2} optimally ranks subsets whose documents are
represented in the same way (e.g., the documents which are indexed by a given group of terms
or share the same set of feature values). In terms of decision, if fallout is fixed, the PRP
permits to decide whether a document (subset) should be retrieved with the minimum probability
of error.
\vfill
\section{Vector Probability}
\label{sec:vect-prob}

When using classical probability term occurrence would correspond to disjoint document subsets
(i.e., a subset corresponds to an index term occurring in every document of the subset). When
using vector probability, term occurrence corresponds to a document vector subspace which is
spanned by the orthonormal vector either $\ket{0}$ or $\ket{1}$ representing, respectively,
absence and presence of a given term. (For the sake of clarity, we consider a single term,
binary weights and binary relevance as depicted in Figure~\ref{fig:ir-decision}.)

As relevance is an event, two vectors represent binary relevance: a relevance vector
$\ket{m_0}$ represents non-relevance state and an orthogonal relevance vector $\ket{m_1}$
represents relevance state.  Relevance vectors and occurrence vectors belong to a common
vector space and thus can be defined in terms of a given orthonormal basis of that space.  

In a vector space, a random variable is a collection of values and of vectors (or
projectors). The vectors are mutually \emph{orthonormal} and 1:1 correspondence with the
values.

Let $x$ be a random variable value (e.g., term occurrence) and $m$ be a conditioning event
(e.g., relevance). In Quantum Theory, $\ket{m}$ is also known as state vector and is a
specialization of a density operator, that is, a Hermitian and unitary trace operator.  The
vector probability that ${x}$ is observed given ${m}$ is $\left|\braket{x}{m}\right|^2$. When
a density operator $\rho$ and an event is represented by projector $\mathbf{P}$, the vector
probability of the event conditioned to the density operator is given by Born's rule, that is,
$\tr(\rho\mathbf{P})$.  When $\rho = \ketbra{m}{m}$ and $\mathbf{P} = \ketbra{x}{x}$, vector
probability is a specialization of Born's rule. (See~\cite{Hughes89}.)

It is possible to show that
\begin{proposition}
  \label{sec:vect-prob-relev}
  A classical probability distribution can be equivalently expressed using vector probability.
\end{proposition}
The proof is in the appendix.

\section{Optimal Vectors}
\label{sec:optim-observ}

In this section, we reformulate the PRP by replacing subsets with vector subspaces, namely, we
replace the notion of optimal document subset with that of optimal vectors (or, vector
subspaces). Such a reformulation allows us to compute the optimal vectors that are more
effective than the optimal document subsets.  To this end, we define a density matrix
representing a probability distribution that has no counterpart in, but that is an extension
of classical probability. Such a density matrix is the outer product of a relevance vector by
itself. When classical probability is assumed, a decision under uncertainty conditions taken
upon this density matrix is equivalent to~\eqref{eq:2} as illustrated in~\cite{Melucci&11c}.
(See the appendix and~\cite{Melucci&11c} as for the details.)  When vector probability is
assumed, a decision under uncertainty conditions taken upon this density matrix is based upon
a different region of acceptance.  Hence, we leverage the following Helstrom's lemma because
it is the rule to compute the optimal vectors.
\begin{lemma}
  \label{the:helstrom}
  Let $\ket{m_1}, \ket{m_0}$ be the relevance vectors.  The optimal vectors $\ket{\mu_0},
  \ket{\mu_1}$ at the highest probability of detection at every probability of false alarm is
  given by the eigenvectors of
  \begin{equation}
    \ketbra{m_1}{m_1} - \lambda\ketbra{m_0}{m_0}
    \label{eq:26}
  \end{equation}
  whose eigenvalues are positive.
\end{lemma}
\begin{proof}
  See~\cite{Helstrom76}.
\end{proof}
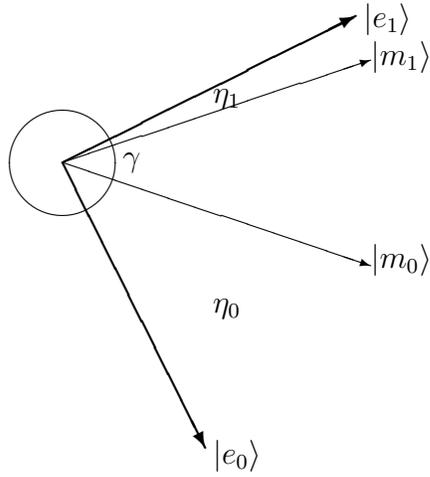
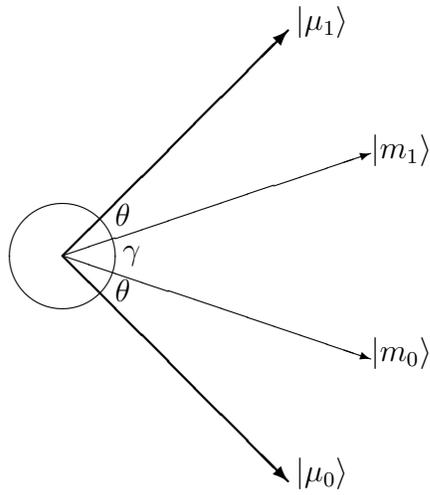
\begin{figure}[t] 
  \centering 
  \setlength{\unitlength}{1mm}  
  \subfigure[Non-optimal  vectors are mutually orthogonal and
  \emph{asymmetrically}  placed around $\ket{m_0}, \ket{m_1}$]{
    \begin{minipage}[t]{1.0\columnwidth}
      \begin{picture}(50,60)(-25,-40)
        \put(0,0){\vector(3,+1){41}}\put(41,13){$\ket{m_1}$}
        \put(0,0){\vector(3,-1){41}}\put(41,-14){$\ket{m_0}$}
        \thicklines
        \put(0,0){\vector(2,+1){39}}\put(40,18){$\ket{e_1}$}
        \put(0,0){\vector(1,-2){19}}\put(20,-40){$\ket{e_0}$}
        \thinlines
        \put(0,0){\circle{100}}
        \put(8,-0.5){$\gamma$}
        \put(20,8){$\eta_1$}
        \put(20,-20){$\eta_0$}
      \end{picture}
   \end{minipage}
    \label{fig:geometry-a}} 
  \subfigure[Optimal  vectors are mutually orthogonal and
  \emph{symmetrically}  placed around $\ket{m_0}, \ket{m_1}$]{
    \begin{minipage}[t]{1.0\columnwidth}
      \begin{picture}(50,65)(-25,-30)
        \put(0,0){\vector(3,+1){41}}\put(41,13){$\ket{m_1}$}
        \put(0,0){\vector(3,-1){41}}\put(41,-14){$\ket{m_0}$}
        \thicklines
        \put(0,0){\vector(1,+1){30}}\put(31,30){$\ket{\mu_1}$}
        \put(0,0){\vector(1,-1){30}}\put(31,-30){$\ket{\mu_0}$}
        \thinlines
        \put(0,0){\circle{100}}
        \put(8,-0.5){$\gamma$}
        \put(7,4){$\theta$}
        \put(7,-6){$\theta$}
      \end{picture}
    \end{minipage}
    \label{fig:geometry-b}}
  \caption{Geometry of decision and  vectors}
  \label{fig:geometry}
\end{figure}

The optimal vectors always exist due to the Spectral Decomposition theorem~\cite{Halmos87};
therefore they are mutually orthogonal because are eigenvectors of~\eqref{eq:26}; moreover,
they can be defined in the space spanned by the relevance vectors.  The angle between the
relevance vectors $\ket{m_1}, \ket{m_0}$ determines the geometry of the decision of the
emitter of Figure~\ref{ir-decision-b}~--~geometry means the probability distributions of the
events.  Therefore, the probability of correct decision and the probability of error are given
by the angle between the two relevance vectors and by the angles between the vectors and the
relevance vectors.  Figure~\ref{fig:geometry-a} depicts the geometry of the optimal
vectors. (The figure is in the two-dimensional space for the sake of clarity, but the reader
should generalize to higher dimensionality than two.)  Suppose $\ket{e_1}, \ket{e_0}$ are
two any other vectors.  The angles $\eta_0, \eta_1$ between the vectors and the relevance
vectors $\ket{m_0}, \ket{m_1}$ are related with the angle $\gamma$ between $\ket{m_0},
\ket{m_1}$ because the vectors are always mutually orthogonal and then the angle is
$\frac{\pi}{2} = \eta_0 + \gamma + \eta_1$.  The optimal vectors are achieved when the angles
between an vector and a relevance vector are equal to
\begin{equation}
  \label{eq:1} 
  \theta = \frac{1}{2}\left(\frac{\pi}{2}-\gamma\right)
\end{equation} 
The rotation of the non-optimal vectors such that~\eqref{eq:1} holds, yields the optimal
vectors $\ket{\mu_1}, \ket{\mu_0}$ as Figure~\ref{fig:geometry-b} illustrates: the optimal
vectors are ``symmetrically'' located around the relevance vectors.  If any two vectors are
rotated in an optimal way, we can achieve the most effective document vector subspaces (or,
vectors) in terms of expected recall and fallout. These vectors cannot be ascribed to the
subsets yielded by dint of the PRP, the latter impossibility being called
incompatibility~\cite{vanRijsbergen04,Hughes89}.

\section{Vector Probability of Relevance}
\label{sec:vect-prob-relev-1}

In this section, we leverage Lemma~\ref{the:helstrom} to introduce the optimal vectors in IR.
We define $\ket{m_0}$ and $\ket{m_1}$ as:
\begin{equation}
  \label{eq:14}
  \ket{m_0} =
  \left(
    \begin{array}{c}
      \sqrt{\poit{1}{m_0}} \\ \sqrt{\poit{0}{m_0}}
    \end{array}
  \right)
  \qquad
  \ket{m_1} = 
  \left(
    \begin{array}{c}
      \sqrt{\poit{1}{m_1}} \\ \sqrt{\poit{0}{m_1}}
    \end{array}
  \right)
\end{equation}
Note that, according to Born's rule $|\braket{m_0}{1}|^2 = \poit{1}{m_0}$ and
$|\braket{m_1}{1}|^2 = \poit{1}{m_1}$, thus \eqref{eq:14} reproduce the classical probability
distributions.  

If the oracle of Figure~\ref{ir-decision-b} exists, an IR system performing like detector B
computes the probabilities that the transformation of a binary symbol referring to an index
term, into $\ket{\mu_0}$ or $\ket{\mu_1}$ occurs in relevant documents and in non-relevant
documents.  The former is called \emph{vector probability of detection} ($Q_d$) and the latter
is called \emph{vector probability of false alarm} ($Q_0$).  These probabilities are the
analogous of $P_0, P_d$. But, if $\ket{\mu_0}, \ket{\mu_1}$ are the mutually exclusive symbols
yielded by the oracle, we have that
\begin{equation}
  \label{eq:3}
  Q_0 = |\braket{m_0}{\mu_1}|^2 \qquad Q_d = |\braket{m_1}{\mu_1}|^2
\end{equation}
The latter expression is \emph{not} the same probability distribution as~\eqref{eq:14} because
it refers to different events.  According to~\cite{Helstrom76}, we have that the \emph{vector
  probability of error} and the \emph{vector probability of correct decision} are defined,
respectively, as
\begin{equation}
  \label{eq:4}
  Q_e = \frac{1}{2}\left(1-\sqrt{1-4\xi(1-\xi)\interm}\right) \qquad Q_c = 1-Q_e
\end{equation}
Both probabilities depend on
\begin{equation}
 \interm = \left(\sqrt{\poit{1}{m_0}\poit{1}{m_1}} + \sqrt{(1-\poit{1}{m_0}) (1-\poit{1}{m_1})}\right)^2
\end{equation}
which is a measure of the distance between two probability distributions as proved
in~\cite{Helstrom76,Wootters81}.  As the probability distributions refer to relevance and
non-relevance, $\interm$ is a measure of the distance between relevance and non-relevance. An
example may be useful.

Suppose, for example, that the probability distributions are $\poit{1}{m_0} = \frac{4}{5}$,
$\poit{1}{m_1} = 1$.  If relevance and non-relevance are equiprobable, $P_e =
\frac{1}{2}\left(P_0 + 1-P_d\right) = \frac{2}{5}$ and $P_c = \frac{1}{2}\left(1-P_0 +
  P_d\right) = \frac{3}{5}$.  When $\lambda=1$, the optimal vectors are the eigenvectors of
\begin{equation*}
  \left(
    \begin{array}{cc}
      \frac{1}{5}  & -\frac{2}{5} \\
      -\frac{2}{5} & -\frac{1}{5}
    \end{array}
  \right)
\end{equation*}
that is,
\begin{equation}
  \label{eq:30}
  \left(\ket{\mu_0}, \ket{\mu_1}\right) = 
  \left(
    \begin{array}{cc}
      -\frac{1}{2}+\frac{\sqrt{5}}{2} & 1 \\
      -\frac{1}{2}-\frac{\sqrt{5}}{2} & 1
    \end{array}
  \right)
  \qquad
\end{equation}
These vectors can be computed in compliance with~\cite{Eldar&01}.  Hence, $Q_e = \frac{1}{10}
\left(5-\sqrt{5}\right)$, which is less than $P_e$. 

The following theorem that is our main result shows that the latter example is not an
exception.
\begin{theorem}
  \label{sec:optim-prob-rank-1}
  If $\poit{x}{m_j}, j=0,1$ are two arbitrary probability distributions conditioned to
  $m_0,m_1$, the latter indicating the probability distribution of term occurrence in
  non-relevant documents and in relevant documents, respectively, then
  \begin{equation}
    \label{eq:23}
    Q_e \leq P_e
  \end{equation}
\end{theorem}
\begin{proof}
  See Section~\ref{sec:proof-prop-refs}.
\end{proof}
Hence, if we were able to find the optimal vectors, retrieval performance of the detector B of
Figure~\ref{fig:geometry-b} would \emph{always} be higher than retrieval performance of the
detector B of Figure~\ref{fig:geometry-a}.

\section{Getting Beyond the State-of-the-Art}
\label{sec:getting-beyond-state}

\begin{figure}[t]
  \centering
  \includegraphics[angle=270]{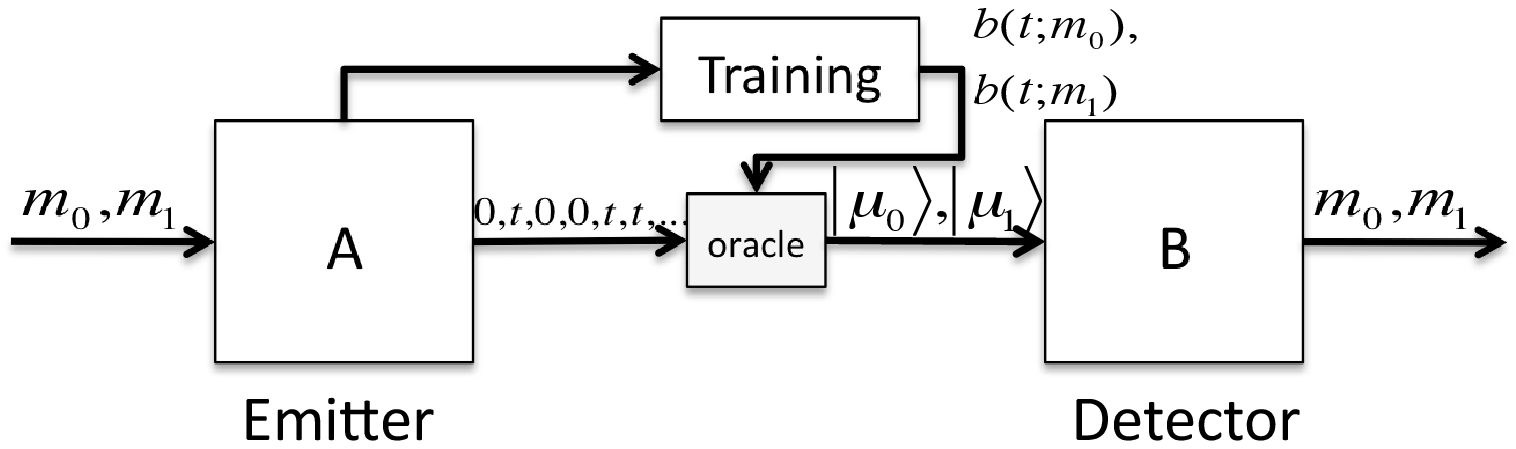}
  \caption{An IR system equipped with an oracle based on vector probability estimated with
    BM25. A document emits the saturation value if the term occurs, $0$ otherwise. The system
    B trains the probability distributions using the saturation values. Rather that applying
    logarithms and computing the BM25 weights, B invokes the oracle that produces the
    vectors. } 
  \label{fig:ir-decision-c}
\end{figure}

The development of the theory assumed binary weights for the sake of clarity. In the event of
non-binary weights, e.g., BM25, we slightly fit the theory as follows. If we do the
development of the BM25 illustrated in~\cite{Robertson&09} in reverse, we can find that the
underlying probability distribution is 
\begin{eqnarray}
 b(t_j;m_i) 	&=& B_{i,j} \poit{1}{m_i}^{t_j}(1-\poit{1}{m_i})^{-t_j} \\ 
  t_j		&=& \mbox{BM25 saturation factor}\\
  B_{i,j}	&=&
  \left\{
    \begin{array}{ll}
      \frac{1}{n} & \poit{1}{m_i} = \frac{1}{2}\\
       \frac{\log\frac{1-\poit{1}{m_i}}{\poit{1}{m_i}}}{1-\left(\frac{\poit{1}{m_i}}{1-\poit{1}{m_i}}\right)^n}
       & \poit{1}{m_i} \neq \frac{1}{2}
    \end{array}
  \right.    \\ 
  n		&=& \max\left\{ t_j \right\}
\end{eqnarray}
thus $b(t_j;m_i)$ is the probability that $t_j$ for term $j$ is observed under state
$m_i$. ($B_{i,j}$ is just a normalization factor.) If $b(t_j;m_i)$ estimates
$\poit{x_j}{m_i}$, the theory still works because Theorem~\ref{sec:optim-prob-rank-1} is
independent of the estimation of $\poit{x_j}{m_i}$. Figure~\ref{fig:ir-decision-c} depicts an
IR system equipped with an oracle based on vector probability estimated with BM25.

\section{Experimental Study}
\label{sec:experiments}

In this section we report an experimental study. Our study differs from the common studies
conducted in usual IR evaluation because: (1) Theorem~\ref{sec:optim-prob-rank-1} already
proves that an IR system working as the detector of Figure~\ref{fig:geometry-b} will always be
more effective than any other system, therefore, if the former were available, every test
would confirm the theorem; (2) as an experimentation that compare two systems using, say, Mean
Average Precision requires the implementation of the oracle, which cannot be at present
implemented, what we can measure is only the degree to which an IR system working as the
detector of Figure~\ref{fig:geometry-b} will outperform any other system.

We have tested the theory illustrated in the previous sections through experiments based on
the TIPSTER test collection, disks 4 and 5. The experiments aimed at measuring the difference
between $P_e$ and $Q_e$ by means of a realistic test collection. To this end, we have used the
TREC-6, 7, 8 topic sets. The queries are topic titles.  We have implemented the following
test: $\poit{x}{m}$ has been computed for each topic, query word and $m \in \{m_0, m_1\}$ by
means of the usual relative frequency of the word within relevant ($m=m_1$) or non-relevant
($m=m_0$) documents. In particular, $x=1$ means presence, $x=0$ means absence.  Thus,
$\poit{1}{m_0}$ is the estimated probability of occurrence in non-relevant documents and
$\poit{1}{m_1}$ is the estimated probability of occurrence in relevant documents. We have
shown in Section~\ref{sec:getting-beyond-state} that the improvement is independent of
probability estimation and then of term weighting.

\begin{figure}[!]
  \centering
  \includegraphics[width=150mm]{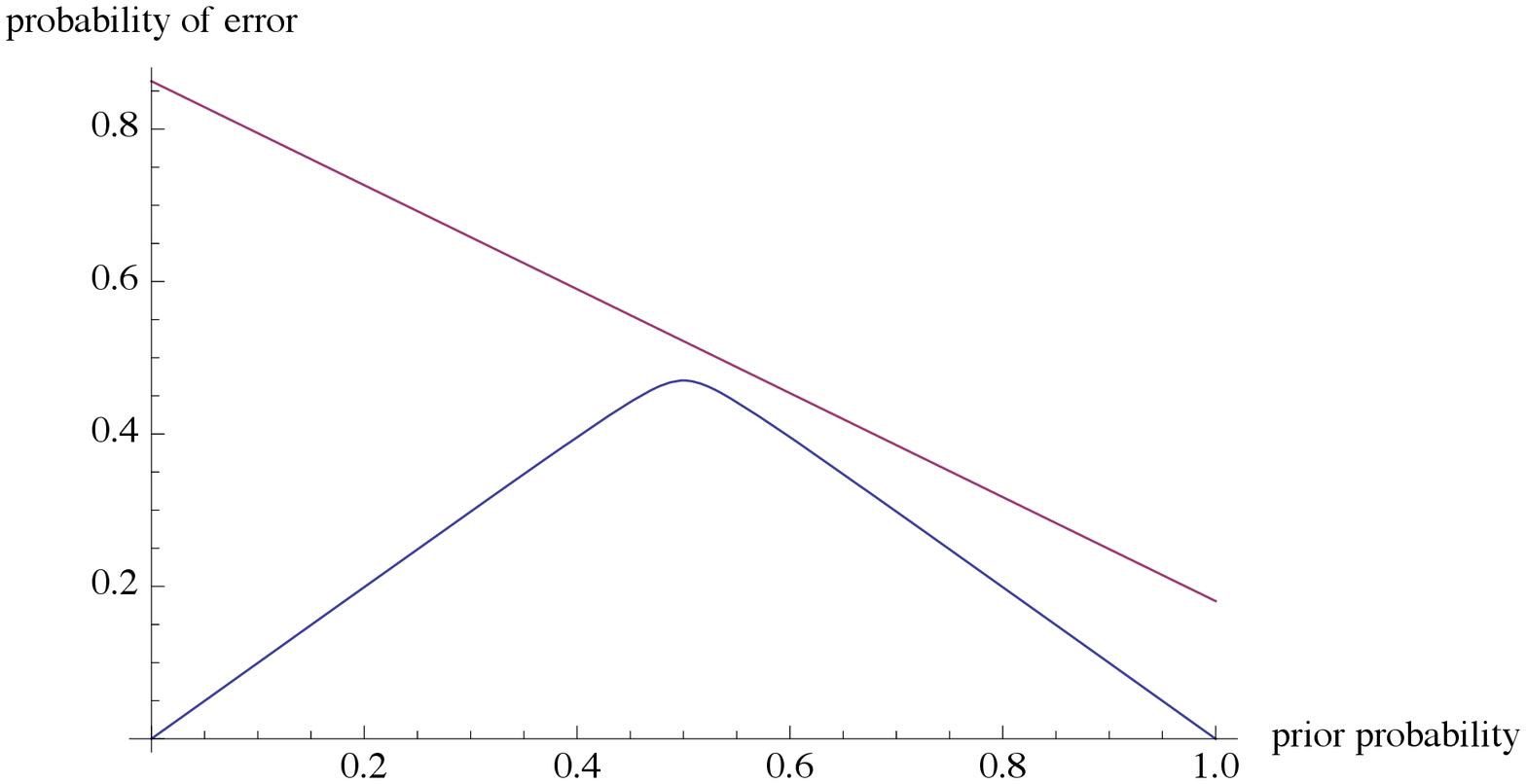}
  \caption{$P_e$ and $Q_e$ plotted against $\xi$ for word \texttt{crime} of topic $301$.}
  \label{fig:t301-crime}
\end{figure}

\begin{figure}[!]
  \centering
  \includegraphics[width=150mm]{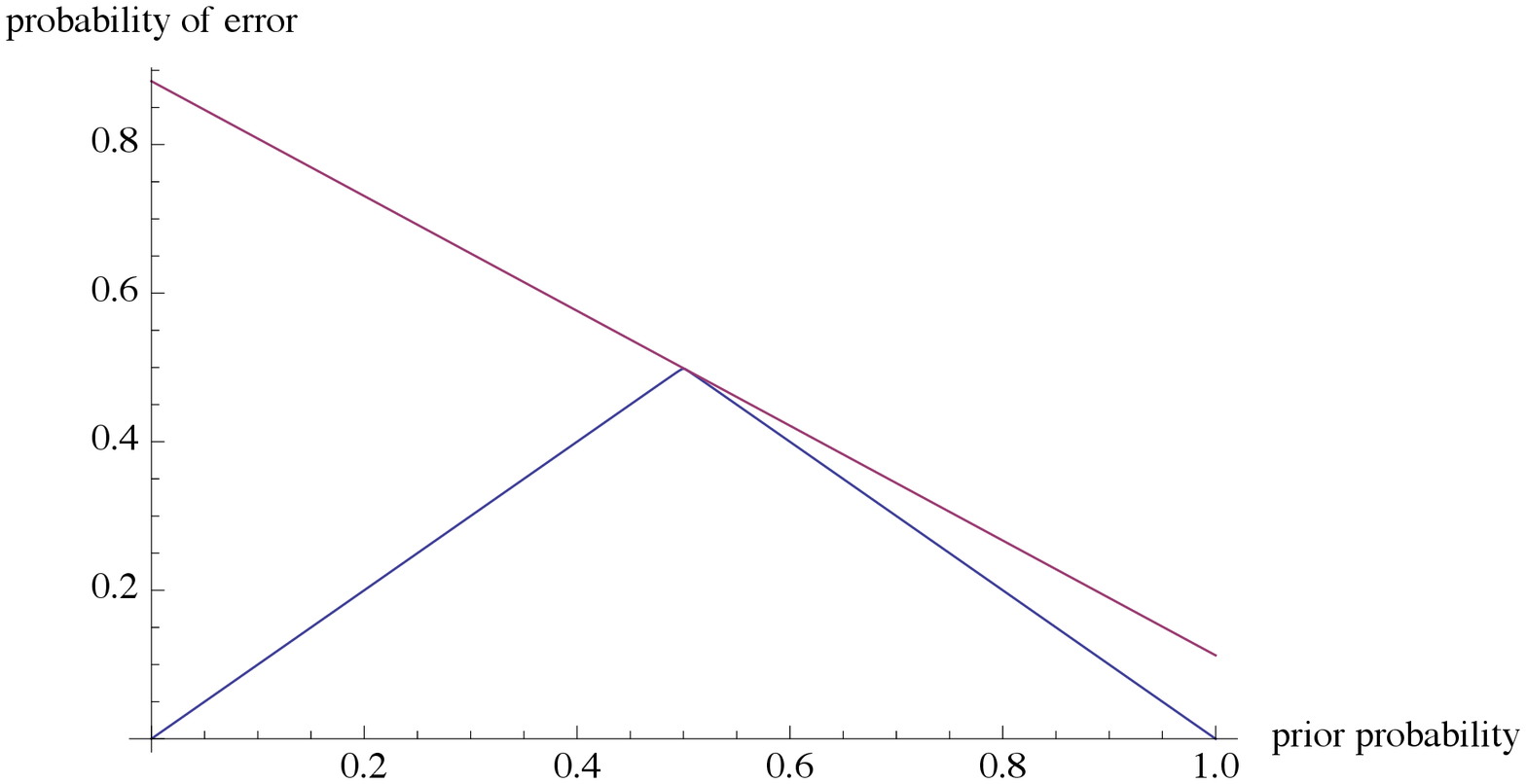}
  \caption{$P_e$ and $Q_e$ plotted against $\xi$ for topic $301$.}
  \label{fig:t301}
\end{figure}

\begin{figure}[!]
  \centering
  \includegraphics[width=150mm]{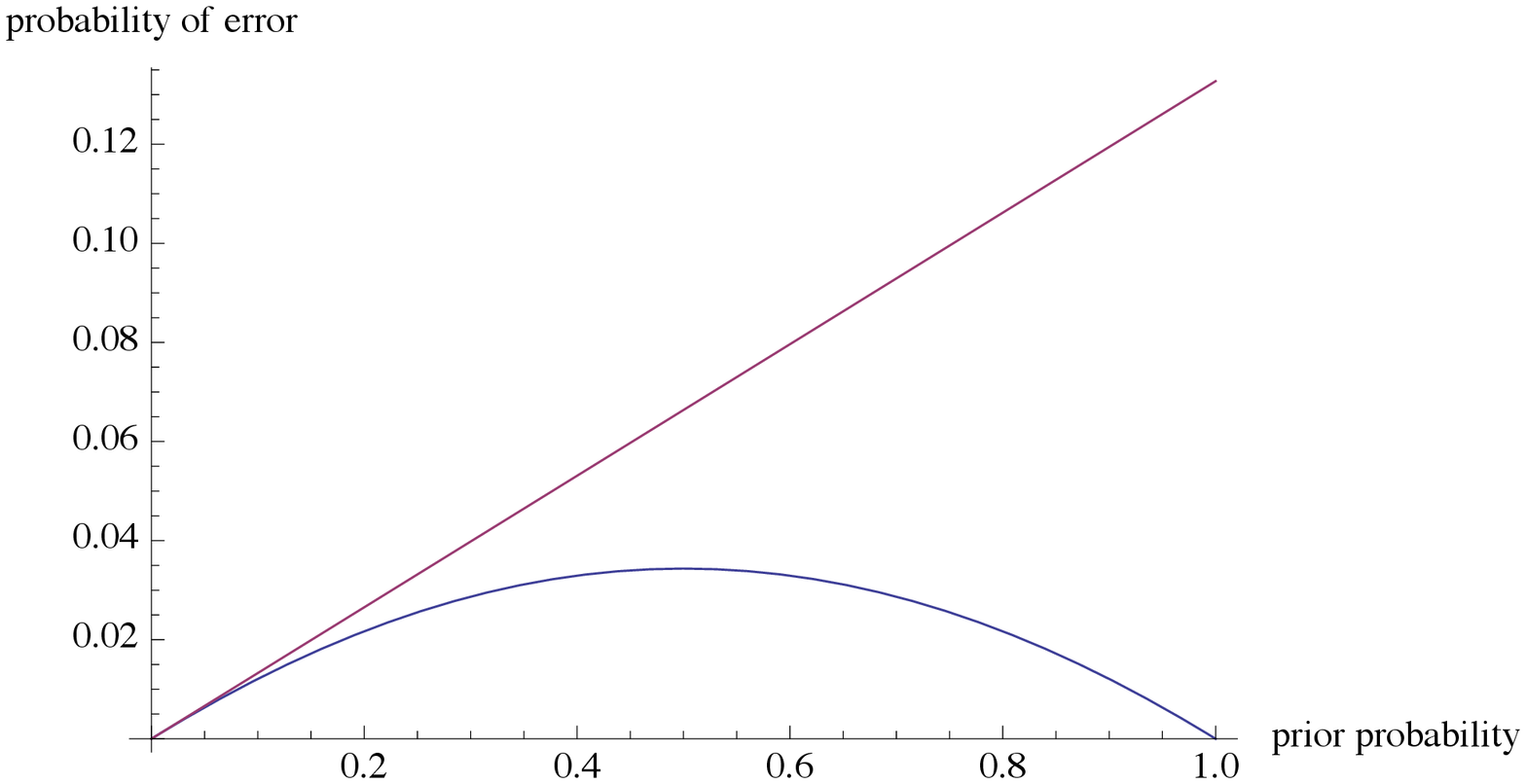}
  \caption{$P_e$ and $Q_e$ plotted against $\xi$ for topic $344$.}
  \label{fig:t344}
\end{figure}

Consider word \texttt{crime} of Topic No. 301; we have that $\poit{1}{m_0} = \frac{223}{1234}$
and $\poit{1}{m_1} = \frac{65}{474}$. Hence, $\interm = 0.998$. (Relevance and non-relevance
probability distributions are very close to each other.)  Estimation has taken advantage of
the availability of the relevance assessments and thus it has been computed on the basis of
the explicit assessments made for each topic.  Figure~\ref{fig:t301-crime} depicts $P_e, Q_e$
as function of the prior probability $\xi$.  $Q_e$ is always greater than $P_e$ for every
prior probability $\xi$. The vertical distance between the curves is due to the value of
$\interm$, which also yields the shape of the $Q_e$ curve, meaning that \texttt{crime}
discriminates between relevant and non-relevant documents to an extent depending on $\interm$
and $\xi$. The average curves computed over all the query words and depicted in
Figure~\ref{fig:t301} give an idea of the overall discriminative power of the topic. In
particular, if the total frequencies within relevant and non-relevant documents are computed
for each query word and a given topic, average probability of error is computed, for each
prior probability. When $\xi$ is close to $\frac{1}{2}$, the curves are indistinguishable
because $\interm$ is very close to $1$. The situation radically changes when Topic No. 344 is
considered because $\interm \approx \frac{9}{25}$; indeed, Figure~\ref{fig:t344} confirms that
$Q_e \gg P_e$ when $\xi \gg 0$.

We have also investigated the event that explicit relevance assessment cannot be used because
of the lack of reliable judgements for a suitable number of documents. In this event, it is
customary to state that $\poit{1}{m_0} = \frac{n+\frac{1}{2}}{N+1}$ and $\poit{1}{m_1} =
\frac{1}{2}$~\cite{Robertson&76}.  Although pseudo-relevance data is assumed, $\interm$ and
$P_e$ can still be computed as function of $\poit{1}{m_0}$ because the latter are still valid
estimations.  In particular, we have that
\begin{equation}
  \label{eq:8}
  \interm = \frac{1}{2}\left(\sqrt{\poit{1}{m_0}} + \sqrt{1-\poit{1}{m_0}}\right)^2
\end{equation}

\begin{figure*}[t]
  \centering
  \includegraphics[width=150mm]{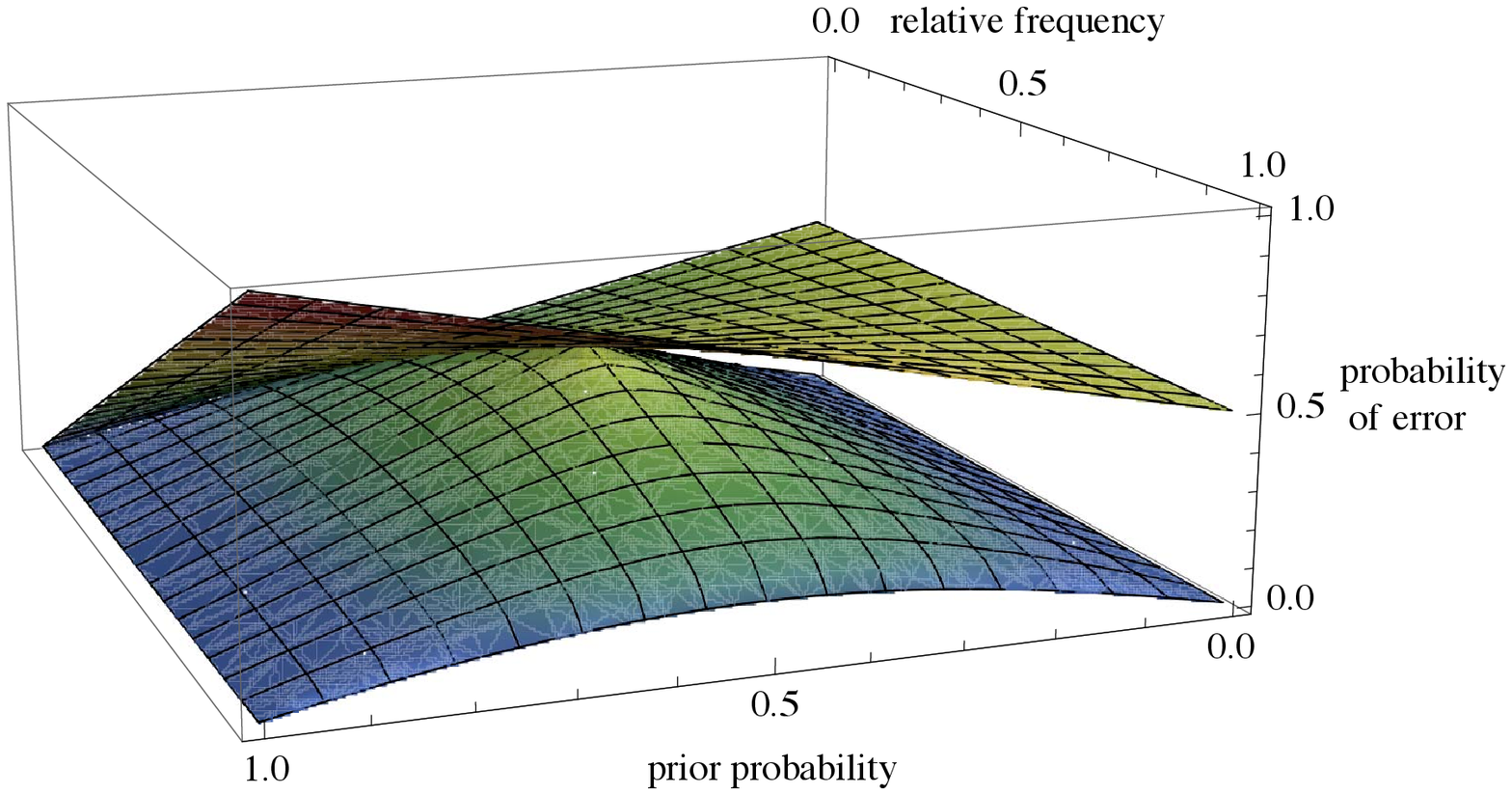}
  \caption{$P_e$ and $Q_e$ plotted against $\xi$ and relative frequency of a term.}
  \label{fig:noqrel}
\end{figure*}

Thus, we can analyze the probabilities of error as functions of $\xi$ and $\poit{1}{m_0}$.
Figure~\ref{fig:noqrel} depicts how $P_e, Q_e$ change with $\xi$ and $\poit{1}{m_0}$; this
plot does not depend on a topic, it rather depends on $\poit{1}{m_0}$, which is a measure of
discrimination power of a query word since $\mathrm{IDF} = -\log \poit{1}{m_0}$. The plot
confirms the intuition that $P_e$ increases when $\poit{1}{m_0}$ increases, that is, when IDF
decreases. In particular, $P_e, Q_e$ are close to each other when little information about the
proportion of relevant documents is available (i.e., $\xi \approx \frac{1}{2}$) and the IDF is
not large enough to make a term discriminative.  Nevertheless, if some information about the
proportion of relevant documents is available (i.e., $\xi$ approaches either $1$ or $0$),
$Q_e$ becomes much smaller than $P_e$ even when the IDF is small (see the bottom-right side of
the plot of Figure~\ref{fig:noqrel}). Table~\ref{tab:avrelfreq-topic} reports the average
relative topic word frequency for each topic computed over the query words. The relative
frequency gives a measure of query difficulty and can be used to ``access'' to the plot of
Figure~\ref{fig:noqrel} to have an idea of $P_e$ when using a classical probabilistic IR model
and of the improvement that can be achieved through an oracle which can produce the optimal
vectors on the basis of the same available evidence as that used to estimate the
$\poit{x}{m}$'s.

\section{Regarding the Oracle}
\label{sec:regarding-oracle}

This section explains why the design of the oracle is difficult.  To this end, the section
refers to some results of logic in IR reported in some detail in~\cite{vanRijsbergen04}.  When
binary term occurrence is considered, there are two mutually exclusive events, i.e., either
presence ($0$) or absence ($1$). The classical probability used in IR is based on
Neyman-Pearson's lemma which states that the set of term occurrences can be partitioned into
two disjoint regions: one region includes all the frequencies such that relevance will be
accepted; the other region denotes rejection\cite{Neyman&33}. If a term is observed from
documents and only presence/absence is observed, the possible regions of acceptance are
$\emptyset, \{0\}, \{1\}, \{0,1\}$.  When using vectors, the regions of acceptance are
$\mathbf{0}, \ketbra{0}{0}, \ketbra{1}{1}, \mathbf{I}$ which are the projectors to,
respectively, the null subspace, the subspace spanned by $\ket{0}$, the subspace spanned by
$\ket{1}$, and the entire space.

Consider the symbols $\ket{\mu_0}, \ket{m_1}$ emitted by the oracle. Vector probability is
based on Lemma~\ref{the:helstrom} which states that the set of symbols can be partitioned into
two disjoint regions: one region includes all the symbols such that relevance will be
accepted; the other region denotes rejection~\cite{Helstrom76}. If a symbol is observed from
the oracle, the regions of acceptance are $\mathbf{0}, \ketbra{\mu_0}{\mu_0},
\ketbra{\mu_1}{\mu_1}, \mathbf{I}$.

\begin{figure}[t]
  \centering
  \includegraphics[angle=270]{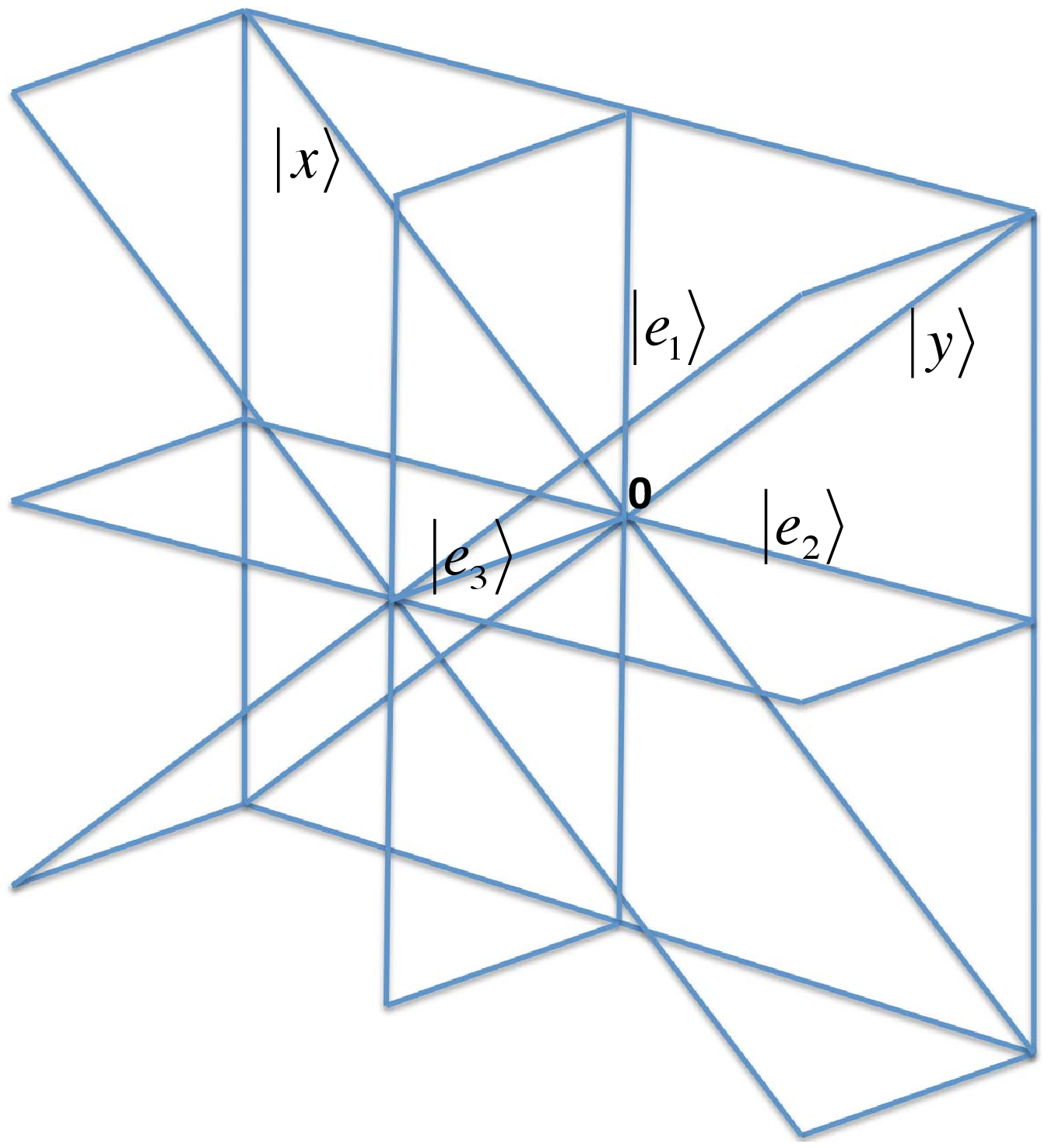}
  \caption{The difference between subsets and vector subspaces. The span of $\ket{y}$ and
    $\ket{x}$ yields the vertical plane, which is spanned by $\ket{e_1}$ and $\ket{e_2}$ too.
    If $\ket{e_2}$ is intersected by the plane, the result is $\ket{e_2}$. But, if $\ket{e_2}$
    is intersected by $\ket{y}$ and then by $\ket{x}$, the span of the two intersections is
    $\mathbf{0}$. That is, the distributive law is not admitted by vectors.}
  \label{fig:subspaces}
\end{figure} 

The problem is that the subspaces spanned by $\ket{\mu_0}, \ket{m_1}$ cannot be defined in
terms of \emph{set} operations on the subspaces spanned by $\ket{0}, \ket{1}$.  Vector
subspaces are equivalent to subsets and they then can be subject to set operations \emph{if}
they are mutually orthogonal~\cite{Griffiths02}. 

To explain the incompatibility between sets and vectors, we illustrate the fact that the
distributive law cannot be admitted in vector spaces as it is in set spaces.
Figure~\ref{fig:subspaces} shows a three-dimensional vector space spanned by $\ket{e_1},
\ket{e_2}, \ket{e_3}$. The ray (i.e.  one-dimensional subspace) $L_x$ is spanned by $\ket{x}$,
the plane (i.e. two-dimensional subspace) $L_{x,y}$ is spanned by $\ket{x}, \ket{y}$. Note
that $L_{e_1,e_2} = L_{x,y} = L_{e_1,y}$ and so on. According to~\cite[page 191]{Hughes89},
consider the vector subspace $L_{e_2} \wedge (L_y \vee L_x)$ provided that $\wedge$ means
``intersection'' and $\vee$ means ``span'' (and not set union). Since $L_y \vee L_x = L_{x,y}
= L_{e_1,e_2}$, $L_{e_2} \wedge (L_y \vee L_x) = L_{e_2} \wedge L_{e_1,e_2} =
L_{e_2}$. However, $(L_{e_2} \wedge L_y) \vee (L_{e_2} \wedge L_x) = 0$ because $L_{e_2}
\wedge L_y = 0$ and $L_{e_2} \wedge L_x = 0$, therefore
\begin{equation} 
  L_{e_2} \wedge (L_y \vee L_x) \neq (L_{e_2} \wedge L_y) \vee (L_{e_2} \wedge L_x)
\end{equation}
thus meaning that the distributive law does not hold, hence, set operations cannot be
applied to vector subspaces.

Incompatibility, that is, the invalidity of the distributive law, is due to the obliquity of
the vectors.  Thus, the optimal vectors cannot be defined in terms of occurrence vectors due
to obliquity and a new ``logic'' must be searched.

The precedent example points out the issue of the measurement of the optimal vectors.
Measurement means the actual finding of the presence / absence of the optimal vectors via an
instrument or device.  The measurement of term occurrence is straightforward because term
occurrence is a physical property measured through an instrument or device. (A program that
reads texts and writes frequencies is sufficient.)  The measurement of the optimal vectors is
much more difficult because to our knowledge any physical property does not correspond to an
optimal vector.  

Despite the difficulty of measuring optimal vectors, one of the main advantages of vector
probability and the results reported in this paper is that the effort to design the oracle for
any other medium than text is comparable to the effort to design the oracle for text because
the limits to observability of the features corresponding to the optimal vectors are actually
those undergone when the informative content of images, video and music must be represented.
Thus, the question is: what should we observe from a document so that the outcome corresponds
to the optimal vector?  The question is not futile because the answer(s) would effect
automatic indexing and retrieval.

\section{Related Work}
\label{sec:related-work}

Van Rijsbergen's book~\cite{vanRijsbergen04} is the point of departure of our work. It
introduced a formalism based on the Hilbert spaces for representing the IR models within a
uniform framework. As the Hilbert spaces have been used for formalizing Quantum Theory, the
book has also suggested the hypothesis that quantum phenomena have their analogues in IR. In
this paper, we are not much interested in investigating whether quantum phenomena have their
analogues in IR, in contrast, we use Hilbert vector spaces for describing probabilistic IR
models and for defining more powerful retrieval functions. 

The latter use of vector spaces in our paper hinges on Helstrom's book~\cite{Helstrom76},
which provides the theoretical foundation for the vector probability and the optimal vectors.
In particular, it deals with optical communication and the detectability of optical signals
and the improvement of the radio frequency-based methods with which their parameters can be
estimated. Within this domain, Helstrom provides the foundations of Quantum Theory for
deciding among alternative probability distributions (e.g. relevance \emph{versus}
non-relevance, in this paper). In this paper, we point to a parallel between signal detection
and relevance detection by corresponding the need to ferret weak signals out of random
background noise to the need to ferret relevance out of term occurrence. Thus, in this paper,
Quantum Theory plays the role of enlarging the horizon of the possible probability
distributions from the classical mixtures used to define classical distributions to quantum
superpositions~\cite{Melucci&11c}, although decision under conditions of uncertainty can still
be treated by the theory of statistical decisions developed by, for example,~\cite{Neyman&33}
and used in IR too.

Eldar and Forney's paper~\cite{Eldar&01} gives an algorithm for computing the optimal vectors
and obtains a new characterization of optimal measurement, and prove that it is optimal in a
least-squares sense.  $\interm$ is the distance between densities defined in~\cite{Wootters81}
and is implemented as the squared cosine of the angle between the subspaces corresponding to
the relevance vectors. The justification of viewing $\interm$ as a distance comes from the
fact that ``the angle in a Hilbert space is the only measure between subspaces, up to a
constant factor, which is invariant under all unitary transformations, that is, under all
possible time evolutions.''~\cite{Wootters81} The latter is the justification given
in~\cite{vanRijsbergen04} of the use of Born's rule for computing what we call vector
probability.

Hughes' book~\cite{Hughes89} is an excellent introduction to Quantum Theory.  In particular,
it addresses incompatibility between observables~--~we have used that explanation to
illustrate the difficulty in implementing the oracle of Figure~\ref{ir-decision-b}. However,
in~\cite{Hughes89}, there is no mention of optimal vectors.  An introduction to quantum
phenomena (i.e., interference, superposition, and entanglement) and Information Retrieval can
be found in~\cite{Melucci&11c}.  In contrast, we do not address quantum phenomena because our
aims is to leverage vector space properties in conjunction with probability. 

In the IR literature, Quantum Theory is receiving more and more interest.
In~\cite{Piwowarski&10b} the authors propose quantum formalism for modeling some IR tasks and
information need aspects.  In contrast, our paper does not limit the research to the
application of an abstract formalism, but exploits the formalism to illustrate how the optimal
vectors significantly improve effectiveness.  In~\cite{Zuccon&10}, the authors propose
$\interm$ for modifying probability of relevance; $\interm$ in conjunction with a cosine of
the angle of a complex number are intended to model quantum correlation (also known as
interference) between relevance assessments. The implementation of interference is left to the
experimenter and that paper provides some suggestions. While~\cite{Zuccon&10} shows that
vector probability induces a different PRP (called Quantum PRP), this paper shows that vector
probability always induces a more powerful ranking than PRP. 

\section{Conclusions}
\label{sec:conclusions}

The research in IR has been traditionally concentrated on extracting and combining evidence as
accurately as possible in the belief that the observed features (e.g., term occurrence, word
frequency) have to ultimately be scalars or structured objects.  The quest for reliable,
effective, efficient retrieval algorithms requires to implement the set of features as best
one can.  The implementation of a set of features is thus an ``answer'' to an implicit
``question'', that is, which is the best \emph{set} of features for achieving effectiveness as
high as possible?  However, the research in IR often yields incremental results, thus arising
the need to achieve an even better answer. To this end, we suggest to ask another
``question'': Which is the best \emph{vector subspace}?

% \bibliographystyle{abbrv} 
% \bibliography{/Users/melo/Documents/Bibliography/general}

\begin{thebibliography}{10}

\bibitem{Eldar&01}
Y.~Eldar and G.~Forney.
\newblock On quantum detection and the square-root measurement.
\newblock {\em IEEE Transactions on Information Theory}, 47(3):858--872, 2001.

\bibitem{Griffiths02}
R.~B. Griffiths.
\newblock {\em Consistent quantum theory}.
\newblock Cambridge University Press, 2002.

\bibitem{Halmos87}
P.~Halmos.
\newblock {\em Finite-dimensional vector spaces}.
\newblock Undergraduate Texts in Mathematics. Springer, 1987.

\bibitem{Helstrom76}
C.~Helstrom.
\newblock {\em Quantum detection and estimation theory}.
\newblock Academic Press, 1976.

\bibitem{Hughes89}
R.~Hughes.
\newblock {\em The structure and interpretation of quantum mechanics}.
\newblock Harvard University Press, 1989.

\bibitem{Kolmogorov56}
A.~Kolmogorov.
\newblock {\em Foundations of the theory of probability}.
\newblock Chelsea Publishing Company, New York, second edition, 1956.

\bibitem{Melucci&11c} 
M.~Melucci and C.~{van Rijsbergen}.
\newblock Quantum mechanics and information retrieval.
\newblock In {\em Advanced Topics in Information Retrieval}. Springer,
  Forthcoming.

\bibitem{Neyman&33}
J.~Neyman and E.~Pearson.
\newblock On the problem of the most efficient tests of statistical hypotheses.
\newblock {\em Philosophical Transactions of the Royal Society, Series A},
  231:289--337, 1933.

\bibitem{Piwowarski&10b}
B.~Piwowarski, I.~Frommholz, M.~Lalmas, and K.~van Rijsbergen.
\newblock What can quantum theory bring to information retrieval.
\newblock In {\em Proceedings of the 19th ACM international conference on
  Information and knowledge management}, CIKM '10, pages 59--68, New York, NY,
  USA, 2010. ACM.

\bibitem{Robertson77}
S.~Robertson.
\newblock The probability ranking principle in information retrieval.
\newblock {\em Journal of Documentation}, 33(4):294--304, 1977.

\bibitem{Robertson&76}
S.~Robertson and K.~{Sparck Jones}.
\newblock Relevance weighting of search terms.
\newblock {\em Journal of the American Society for Information Science},
  27:129--146, May 1976.

\bibitem{Robertson&09}
S.~Robertson and H.~Zaragoza.
\newblock The probabilistic relevance framework: {BM25} and beyond.
\newblock {\em Foundations and Trends in Information Retrieval}, 3(4):333--389,
  2009.

\bibitem{vanRijsbergen04}
K.~{van Rijsbergen}.
\newblock {\em The geometry of information retrieval}.
\newblock Cambridge University Press, UK, 2004.

\bibitem{Wootters81}
W.~K. Wootters.
\newblock Statistical distance and {H}ilbert space.
\newblock {\em Phys. Rev. D}, 23(2):357--362, Jan 1981.

\bibitem{Zuccon&10}
G.~Zuccon and L.~Azzopardi.
\newblock Using the quantum probability ranking principle to rank
  interdependent documents.
\newblock In {\em Proceedings of the European Conference on Information
  Retrieval Research (ECIR)}, pages 357--369, 2010.

\end{thebibliography}

\appendix{ }
\label{sec:appendix}

\paragraph{Definitions and concepts}
\label{sec:definitions}

% \begin{definition}[Observable]
%   An observable is a property that can be measured from an entity.
% \end{definition}
% Term frequency or relevance are observables.
\begin{definition}[Probability Distribution]
  A probability distribution maps observable values to the real range $\left[0, 1\right]$. As
  usual, the probabilities are not negative and sums to $1$.
\end{definition}
\begin{definition}[Classical Probability Distribution]
  A classical probability distribution admits only \emph{sets} of values.
\end{definition}
The subsets of values can be defined by means of the set operations (i.e., intersection,
union, complement). Thus, one can compute, for instance, the set of relevant documents with a
given term frequency.
% \begin{definition}[State]
%   A state, or hypothesis, is a condition of the measured entity and molds the probability
%   distribution of the measurement. 
% \end{definition}
% In classical probability, ``hypothesis'' is more used than ``state''.  We use ``state''
% because it is used in the formalism of Quantum Theory. We correspond the null state to
% non-relevance and the alternative state to relevance. An IR system decides between the
% relevance state and the non-relevance provided an observable value.
% \begin{definition}[Superposition]
%   In Physics, superposition models observables that are known only if they are measured.  In
%   IR, the event that an observable exists only if observed is a much debated hypothesis.
% \end{definition}
\begin{definition}[Probability of Detection]
  It is the probability that a detector decides for relevance when relevance is true; it is
  called (expected) recall in IR.
\end{definition}
\begin{definition}[Probability of False Alarm]
  It is the probability that a detector decides for relevance when relevance is false; it is
  called (expected) fallout in IR.
\end{definition}
\begin{definition}[Region of Acceptance]
  It is the set of the observable values that induce the system to decide for relevance. The
  most powerful region of acceptance yields the maximum probability of detection for a fixed
  probability of false alarm.
\end{definition}
For example, a region of acceptance is a set of term frequencies.  The Neyman-Pearson lemma
states that the maximum likelihood ratio test defines the most powerful region of
acceptance~\cite{Neyman&33}.
\begin{definition}[Probability of Correct Decision]
  \begin{equation}
    \label{eq:18}
    P_c = \xi (1-P_0) + (1-\xi)P_d
  \end{equation}
  provided that $\xi$ the prior probability of non-relevance, $P_0$ is the probability of
  false alarm and $P_d$ is the probability of detection.
\end{definition}
\begin{definition}[Probability of Error]
  \begin{equation}
    \label{eq:20}
    P_e = \xi P_0  + (1-\xi) (1-P_d)
  \end{equation}
\end{definition}
Of course, $P_e+P_c=1$.  In the following, we adopt the Dirac notation to write vectors so
that the reader may refer to the literature on Quantum Theory; a brief illustration of the
Dirac notation is in~\cite{vanRijsbergen04}.
\begin{definition}[Vector Space]
  A vector space over a field $\cal F$ is a set of vectors subject to linearity, namely, a set
  such that, for every vector $\ket{u}$, there are three scalars $a,b,c \in \cal F$ and three
  vectors $\ket{v}, \ket{x}, \ket{y}$ of the same space such that $\ket{u} = a\ket{v}$ and
  $\ket{u} = b\ket{x}+c\ket{y}$. If $\ket{u}$ is a vector, $\bra{u}$ is its transpose,
  $\braket{v}{u}$ is the \emph{inner product} with $\ket{v}$ and $\ketbra{v}{u}$ is the
  \emph{outer product} with $\ket{v}$. A projector $\mathbf{P}$ is a linear operator acting on
  a vector space such that $\mathbf{P}^n = \mathbf{P}$ for every $n > 0$. In particular,
  $\ketbra{u}{u}$ is the \emph{projector} to the subspace spanned by $\ket{u}$.  If
  $\left|\braket{x}{x}\right|^2 = 1$, the vector is normal. If $\braket{x}{y} = 0$, the
  vectors are mutually orthogonal. A subspace is a \emph{span} of one or more subspaces if its
  projector is a linear combination of the projectors of the latter; for example, a ray is a
  span of a vector, a plane is a span of two rays (or vectors), and so on.
\end{definition}
\begin{definition}[Random Variable]
  In classical probability, a random variable is a collection of values and of sets. The sets
  are mutually \emph{disjoint} and 1:1 correspondence with the values. 
\end{definition}

\paragraph{Proof of Proposition~\ref{sec:vect-prob-relev}}
\label{sec:proof-proposition}

Suppose that $\poit{x}{m}$ is the probability that frequency $x$ is observed given a parameter
$m$ corresponding to relevance. Note that $m$ may refer to more than one parameter. However,
we assume that $m$ is scalar for the sake of clarity.  In the event of binary relevance, $m$
is either $m_0$ (non-relevance) or $m_1$ (relevance).  The expressions
\begin{eqnarray}
  \ket{m_0} = \sum_{x=0}^{N} \amp{x}{m_0} \ket{x} \nonumber\\ 
  \ket{m_1} = \sum_{x=0}^{N} \amp{x}{m_1} \ket{x} \nonumber\\
  \amp{x}{m} = \pm \sqrt{\poit{x}{m}}
  \label{eq:5} 
\end{eqnarray}
establish the relationship between classical probability distributions and vector probability,
namely, between the parameters $m_0, m_1$, the relevance vectors $\ket{m_0}, \ket{m_1}$ and
the observable $X$.  The sign of $\amp{x}{m}$ is chosen so that the orthogonality between the
relevance vectors is retained.  Moreover, the orthogonality of the relevance vectors and
the following expression
\begin{eqnarray}
  \label{eq:9}
  \left|\braket{y}{m}\right|^2&=& \left|\sum_{x=0}^N \amp{x}{m}  \braket{y}{x}\right|^2\\
  {}			      &=& \left|\amp{y}{m}\right|^2 \qquad \mbox{due to orthogonality}\\
  {}			      &=& \poit{y}{m}
\end{eqnarray}
establish the relationship between classical and vector probability of relevance.

\paragraph{Proof of Theorem~\ref{sec:optim-prob-rank-1}}
\label{sec:proof-prop-refs}
 Consider Figures~\ref{fig:geometry-a} and~\ref{fig:geometry-b}.  A probability of
  detection $p_d$ and a probability of false alarm $p_0$ defines the coordinates of
  $\ket{m_0}$ and $\ket{m_1}$ with a given orthonormal basis $\ket{e_0}, \ket{e_1}$ (that
  is, an observable):
  \begin{eqnarray}
    \ket{m_0} = \sqrt{1-p_d}\ket{e_0}+\sqrt{p_d}\ket{e_1}
    \\
    \ket{m_1} = \sqrt{p_0}\ket{e_0}+\sqrt{1-p_0}\ket{e_1}
    \label{eq:22}
  \end{eqnarray}
  The coordinates are expressed in terms of angles:
  \begin{equation}
    1-p_d=\sin^2\eta_1 \qquad p_0 = \sin^2\eta_0
  \end{equation}
  provided that $\eta_i$ is the angle between $\ket{m_i}$ and $\ket{e_1}$.  

  The probability of error is
  \begin{equation}
    p_e = \xi p_0 + (1-\xi)(1-p_d) = \xi \sin^2\eta_0 + (1-\xi)\sin^2\eta_1
  \end{equation}
  The probability of error is minimum when $\eta_0=\eta_1=\theta$ as shown in~\cite[page
  99]{Helstrom76}.

  But, $\theta$ is exactly the angle between $\ket{m_i}, i=0,1$ and $\ket{\mu_i}$ and is
  defined as a result of Equation~\eqref{eq:1}.  The probability of error is then minimized
  when the observable vectors are the $\ket{\mu_i}, i=0,1$.

  Therefore, $Q_e \leq P_e$ for all $P_e$, that is, for all the observable vectors. As
  $Q_c = 1-Q_e$, the probability of correct decision is also maximum.
\vfill
\begin{table*}[p]
 \footnotesize
  \begin{tabular}[t]{|l|r||l|r||l|r|}
    \hline
    Topic  &Av. Relative Frequency&Topic &Av. Relative Frequency&Topic&Av. Relative Frequency\\
    \hline
    301    &             0.0458  &   351    &             0.0054& 401    &             0.0666\\
    302    &             0.0063  &   352    &             0.0284& 402    &             0.0001\\
    303    &             0.0008  &   353    &             0.0018& 403    &             0.0000\\
    304    &             0.0026  &   354    &             0.0016& 404    &             0.0458\\
    305    &             0.0054  &   355    &             0.0017& 405    &             0.0025\\
    306    &             0.0088  &   356    &             0.0107& 406    &             0.0025\\
    307    &             0.0074  &   357    &             0.0055& 407    &             0.0012\\
    308    &             0.0001  &   358    &             0.0033& 408    &             0.0005\\
    309    &             0.0052  &   359    &             0.0131& 409    &             0.0092\\
    310    &             0.0104  &   360    &             0.0164& 410    &             0.0260\\
    311    &             0.0128  &   361    &             0.0016& 411    &             0.0001\\
    312	   &		 0.0000  &   362    &             0.0085& 412    &             0.0343\\
    313    &             0.0003  &   363    &             0.0055& 413    &             0.0251\\
    314    &             0.0026  &   364    &             0.0001& 414    &             0.0123\\
    315    &             0.0029  &   365    &             0.0053& 415    &             0.0070\\
    316    &	         0.0000  &   366    &             0.0108& 416    &             0.0130\\
    317    &             0.0001  &   367    &             0.0002& 417    &             0.0001\\
    318    &             0.0379  &   368    &             0.0001& 418    &             0.0148\\
    319    &             0.0124  &   369    &             0.0000& 419    &             0.0009\\
    320    &             0.0045  &   370    &             0.0266& 420    &             0.0012\\
    321    &             0.0112  &   371    &             0.0272& 421    &             0.0189\\
    322    &             0.0510  &   372    &             0.0218& 422    &             0.0072\\
    323    &             0.0005  &   373    &             0.0162& 423    &             0.0008\\
    324    &             0.0421  &   374    &             0.0021& 424    &             0.0000\\
    325    &             0.0002  &   375    &             0.0082& 425    &             0.0193\\
    326    &             0.0007  &   376    &             0.0571& 426    &             0.0225\\
    327    &             0.0020  &   377    &             0.0010& 427    &             0.0035\\
    328    &             0.0007  &   378    &             0.0103& 428    &             0.0203\\
    329    &             0.0217  &   379    &             0.0000& 429    &             0.0023\\
    330    &             0.0236  &   380    &             0.0103& 430    &             0.0034\\
    331    &             0.0808  &   381    &             0.0026& 431    &             0.0094\\
    332    &             0.0495  &   382    &             0.0059& 432    &             0.0278\\
    333    &             0.0026  &   383    &             0.0067& 433    &             0.0039\\
    334    &             0.0077  &   384    &             0.0119& 434    &             0.0188\\
    335    &             0.0042  &   385    &             0.0110& 435    &             0.0226\\
    336    &             0.0126  &   386    &             0.0134& 436    &             0.0023\\
    337    &             0.0003  &   387    &             0.0060& 437    &             0.0150\\
    338    &             0.0040  &   388    &             0.0015& 438    &             0.0203\\
    339    &             0.0161  &   389    &             0.0144& 439    &             0.0024\\
    340    &             0.0143  &   390    &             0.0054& 440    &             0.0087\\
    341    &             0.0343  &   391    &             0.0340& 441    &             0.0024\\
    342    &             0.0031  &   392    &             0.0000& 442    &             0.0014\\
    343    &             0.0282  &   393    &             0.0021& 443    &             0.0345\\
    344    &             0.0023  &   394    &             0.0599& 444    &             0.0001\\
    345    &             0.0336  &   395    &             0.0014& 445    &             0.0111\\
    346    &             0.0068  &   396    &             0.0157& 446    &             0.0054\\
    347    &             0.0012  &   397    &             0.0008& 447    &             0.0026\\
    348    &             0.0000  &   398    &             0.0199& 448    &             0.0146\\
    349    &             0.0000  &   399    &             0.0012& 449    &             0.0001\\
    350    &             0.0229  &   400    &             0.0035& 450    &             0.0256\\
   \hline
  \end{tabular}
  \caption{Average relative frequency per topic.}
  \label{tab:avrelfreq-topic}
\end{table*}

\end{document}